\documentclass[aps,superscriptaddress,prb,reprint,showpacs]{revtex4-1}
\usepackage{graphicx}
\usepackage{amssymb}
\begin{document}

\title{Atomistic quantum transport modeling of metal-graphene nanoribbon heterojunctions}

\author{I. Deretzis}
\email{ioannis.deretzis@imm.cnr.it}
\affiliation{Scuola Superiore, Universit\`{a} di Catania, I-95123 Catania, Italy}
\affiliation{CNR-IMM, I-95121 Catania, Italy}

\author{G. Fiori}
\affiliation{Dipartimento di Ingegneria dell'Informazione: Elettronica, Informatica, Telecomunicazioni, Universit\`{a} di Pisa, I-56122 Pisa, Italy}

\author{G. Iannaccone}
\affiliation{Dipartimento di Ingegneria dell'Informazione: Elettronica, Informatica, Telecomunicazioni, Universit\`{a} di Pisa, I-56122 Pisa, Italy}

\author{A. La Magna}
\affiliation{CNR-IMM, I-95121 Catania, Italy}
\date{\today}

\begin{abstract}
We calculate quantum transport for metal-graphene nanoribbon heterojunctions within the atomistic self-consistent Schr{\"o}dinger/Poisson scheme. Attention is paid on both the chemical aspects of the interface bonding as well the one-dimensional electrostatics along the ribbon length. Band-bending and doping effects strongly influence the transport properties, giving rise to conductance asymmetries and a selective suppression of the subband formation. Junction electrostatics and $p$-type characteristics drive the conduction mechanism in the case of high work function $Au$, $Pd$ and $Pt$ electrodes, while contact resistance becomes dominant in the case of $Al$.
\end{abstract}

\maketitle

Integration of graphene-based nanostructures in electronics, sensors and environmental applications makes necessary a clear understanding of the interaction between graphene and metallic surfaces\cite{2009NanoL...9..422L,2009PhRvB..79x5430M,2008NatNa...3..486L,2008PhRvB..78p1409L}. Interface bonding and electrostatics can play a crucial role in the transport characteristics of these systems since the low-dimensionality and high carrier mobility of the channel material\cite{2007NatMa...6..183G} can enhance the role of the metallic contact with respect to the traditional complementary metal-oxide semiconductor technology. In this sense it can be argued that the main source of resistivity in graphene-based devices should derive from the interaction with the metallic electrodes. Characteristics of such interaction for two-dimensional graphene have been identified both experimentally\cite{2009PhRvB..79x5430M,2008NatNa...3..486L,2009SSCom.149.1068B} and theoretically\cite{2009PhRvB..79s5425K,2010PhRvL.104g6807B}, where charge transfer, doping-related phenomena and near-interface potential fluctuations have been reported. However, as patterning and lithographic techniques advance towards one-dimensional (1D) confinement in order to engineer the necessary bandgaps for digital applications, a particular 1D electrostatic response can be expected that should strongly differentiate device characteristics with respect to the two-dimensional case\cite{1999PhRvL..83.5174L}. Under this perspective we study metal-graphene nanoribbon (GNR) heterostructures within self-consistent quantum transport simulations on the basis of: a) an atomistic description on both the active device part and the metallic electrode that respects the interface chemical bonding, b) a proper treatment of the junction electrostatics and c) depletion region length-scales. Results show that band-bending and doping effects can significantly alter the ideal transport characteristics of GNRs giving rise to asymmetries in the conductance and a selective suppression of the 1D subband formation. Moreover, electrode-dependent scattering processes can block conduction channels in particular cases. Similar to carbon nanotubes (CNTs), we find that there are long-range depletion tails in the charge distribution\cite{1999PhRvL..83.5174L} that vary on the basis of the conductive character of the respective GNR\cite{2008JChPh.128p4706D}. 

\begin{figure}
	\centering
		\includegraphics[width=0.7\columnwidth]{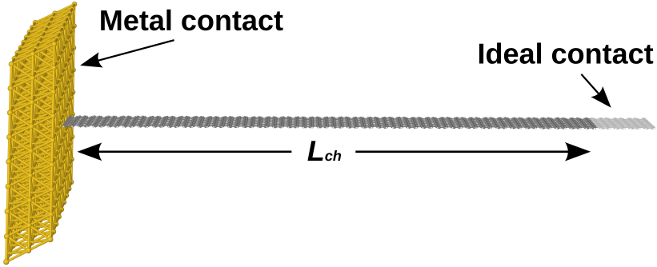}
	\caption{Configuration scheme of the simulated graphene nanoribbon systems, where a GNR is end-contacted by a three-dimensional semi-infinite metallic electrode at the left side, whereas ideally contacted at the right side.}
	\label{fig:geometry}
\end{figure}

We consider hydrogen-terminated armchair and zigzag graphene nanoribbons (aGNRs and zGNRs respectively) and use the terminology of Ref. \citenum{2006PhRvL..97u6803S} to categorize them on the basis of the dimer lines $N_a$ (zigzag chains $N_z$) along the ribbon width. Fig. \ref{fig:geometry} shows the two-terminal geometry used throughout this study, where GNRs with channel lengths $L_{ch} \approx 17nm$ are end-contacted at the left side by the $(111)$ surfaces of three-dimensional semi-infinite electrodes ($Au$, $Pd$, $Pt$ and $Al$). The right electrode is an ideal ohmic contact\cite{2006ITED...53.1782F}, i.e. a GNR with the same dimer lines (zigzag chains) as the device part. In the case of semiconducting aGNRs this geometry corresponds to a Schottky junction. We employ a self-consistent Schr{\"o}dinger/Poisson scheme for the calculation of transport and electrostatics. Quantum transport is computed within the non equilibrium Green's function formalism (NEGF) coupled to the standard Landauer-Buttiker approach\cite{Datta}: the single particle retarded Green's function matrix reads $\mathcal{G} = [E S - H - \Sigma_L - \Sigma_R]^{-1}$, where $E$ is the energy, $H$ ($S$) is the device Hamiltonian (overlap) matrix and $\Sigma_{L,R}$ are self-energies that account for the effect of scattering by the contacts ($\Sigma = {\tau}g_s{\tau}^\dagger$, where $g_s$ is the surface Green function specific to the contact type and $\tau$ is the Hamiltonian relative to the interaction between the device and the contact). From the total transmission probability $T=Trace[{\Gamma_L}\mathcal{G}{\Gamma_R}\mathcal{G}^\dagger]$ (where $\Gamma_{L,R}=i[\Sigma_{L,R}-\Sigma_{L,R}^\dagger]$) conductance can be calculated as $G=(2e^2/h)T$. The device spectral function is the anti-hermitian part of the Green matrix $A=i(\mathcal{G}-\mathcal{G}^{\dagger})$, from which the local density of states (LDOS) at energy $E$ and position $\mathbf{r_{\alpha}}$ can be defined as: $ LDOS(\mathbf{r_{\alpha}},E) = \int_{\mathbf{R^3}}  Trace [ AS/(2 \pi) ] \delta(\mathbf{r}-\mathbf{r_{\alpha}}) d\mathbf{r} $, where $\delta$ is the Delta function and $\mathbf{r_{\alpha}}$ shows the positions of the atomic sites. Hamiltonian and overlap matrices are written within a first-principles-based parameterized model using the extended H{\"u}ckel theory\cite{2000PhRvB..61.7965C,2006JAP...100d3714K} and a non-orthogonal double-$\zeta$ Slater-type basis that fits the bandstructure of bulk graphene\cite{2006JAP...100d3714K} and fcc metals\cite{2000PhRvB..61.7965C} from density functional theory calculations. Metal surface Green functions for the evaluation of the respective self-energies are calculated for the three-dimensional semi-infinite contact with a back-and-forth real to $k$-space Fourier transform exploiting lattice periodicity\cite{Zahi03}. Charging effects are introduced in the formalism with the inclusion of a self-consistent potential $U_{sc}(\rho_f)$ that is a functional of the device density matrix and is added to the bare device Hamiltonian. Within the self-consistent procedure, mobile charges $\rho_f$ deriving from the NEGF are passed to a three-dimensional numerical Poisson solver $\nabla^2 U_{sc}=-\rho_f/\epsilon$, considering the device part embedded in $SiO_2$\cite{2006ITED...53.1782F}. A Dirichlet boundary condition is set in the metal-GNR interface of the Poisson box with a value $U_{sc}^{left}=\phi_{m}-\phi_{gr}$, where $\phi_{m}$, $\phi_{gr}$ are the experimentally measured work functions for (111) metallic surfaces and graphene\footnote{$\phi_{Au_{(111)}}=5.31eV\cite{1977JAP....48.4729M}, \phi_{Pd_{(111)}}=5.6eV\cite{1977JAP....48.4729M}, \phi_{Pt_{(111)}}=5.7eV\cite{1977JAP....48.4729M}, \phi_{Al_{(111)}}=4.24eV$\cite{1977JAP....48.4729M} and $\phi_{gr}=4.6eV$\cite{1997JPCM....9....1O}}. Null Neumann boundary conditions are set for the other five faces of the Poisson simulation box. Self-consistency is enhanced by a predictor/corrector Newton-Rapson algorithm\cite{1997JAP....81.7880T} while optimized matrix manipulation techniques\cite{Petersen20095020} have been implemented throughout the numerical code. Fermi-Dirac statistics are introduced for room temperatures ($300 K$).

\begin{figure}
	\centering
		\includegraphics[width=0.95\columnwidth]{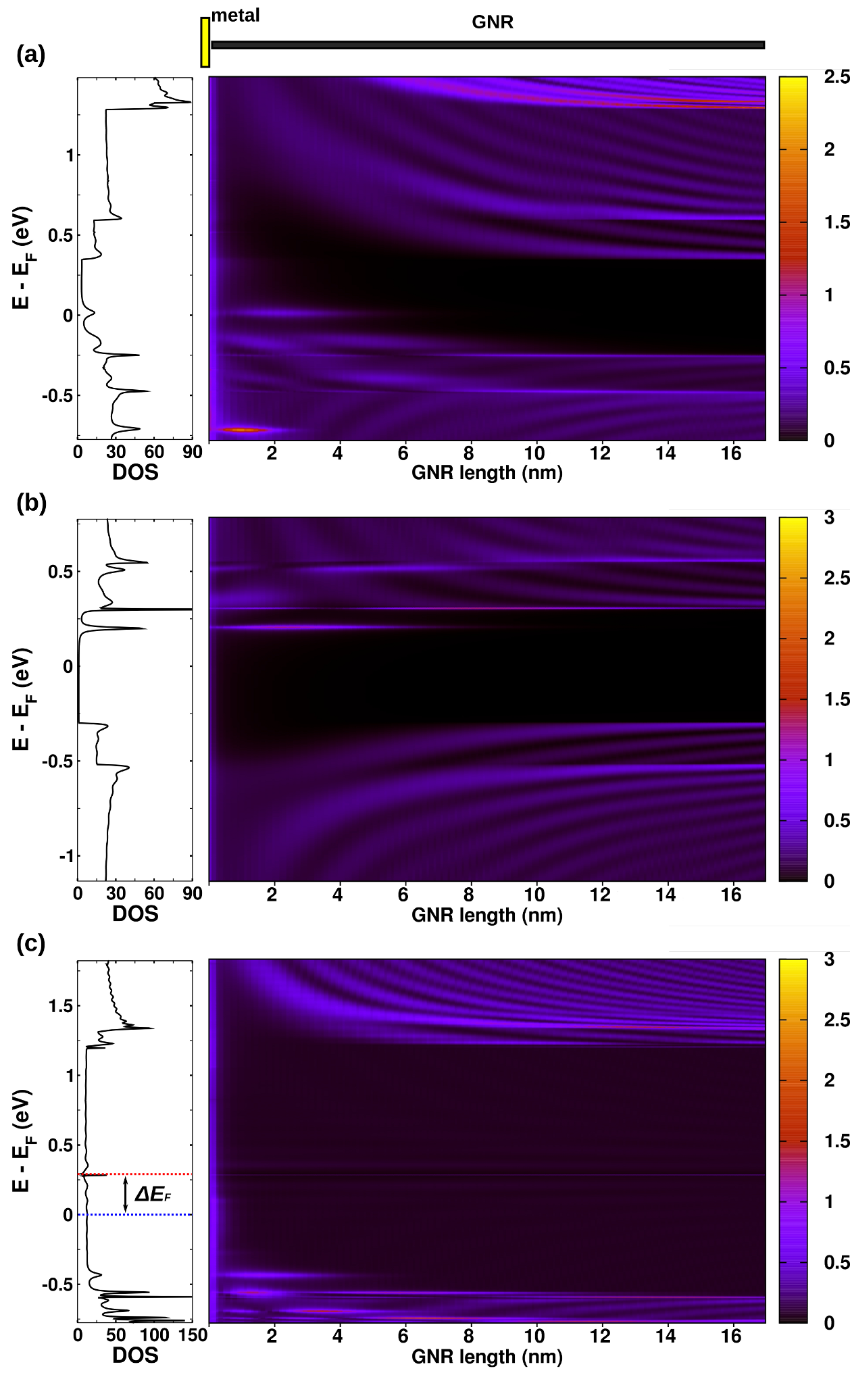}
	\caption{Density of States (left) and real-space bands along the GNR length (right) for (a) a $N_a$=16 aGNR contacted with $Au$, (b) a $N_a$=16 aGNR contacted with $Al$ and (c) a $N_a$=14 aGNR contacted with $Pt$. $\Delta E_F$ in (c) denotes the difference between the Fermi levels of the metal-contacted and the respective ideal aGNR.}
	\label{fig:ldos}
\end{figure}

Fig. \ref{fig:ldos} shows a real-space representation of the band formation along the ribbon lengths within total/local density of states spectra for a semiconducting $N_a$=16 aGNR and a semimetallic $N_a$=14 aGNR. In the case of the 16 aGNR contacted with the high work function $Au$ electrode (Fig. \ref{fig:ldos}(a)) the equilibrium Fermi level alignment for the two parts of the heterostructure gives rise to significant upwards band-bending phenomena near the metal-aGNR interface due to the higher work function of the metal with respect to the GNR. However, band-bending is not rigid for both conduction and valence bands as a result of a complex interference mechanism: the LDOS distribution clearly shows the presence of wavelike quantum interference patterns due to the reflection of the incident electron wave by the non-ideal contact\cite{2003cond.mat.12551G}. Near the interface such patterns tend to turn upwards for the conduction band and downwards for the valence band and respond differently in the presence of the electric field induced by the barrier. Hence, conduction band shifts smoothly while valence band shows localization patterns in the LDOS distribution. Such patterns become discrete localized states with a few-$nm$ spatial breadth in the energy region where the bended valence band is triangularly-like confined inside the bandgap. In addition, metal-induced gap states (MIGS), i.e. tails of the metallic wavefunctions decaying very fast in the semiconducting gap, form throughout the interface (visible as a brighter left-border line for all energies in the LDOS representation of Fig. \ref{fig:ldos}(a)). It can be therefore argued that the interface between a GNR and a metallic contact is ruled by complex band-bending, interference and localization phenomena whose influence in the conduction mechanism will be discussed in the following. When the same aGNR is contacted by a low-work function $Al$ electrode (Fig. \ref{fig:ldos}(b)) the bands bend downwards ($\phi_{Al}-\phi_{GNR}<0$ here), whereas qualitatively similar behaviors as before (interference patterns, localized gap states, MIGS) can be observed. In both cases the Fermi level remains within the bandgap although loosing the midgap position of the respective ideal aGNR. In the case of a semimetallic 14 aGNR contacted with $Pt$ (Fig. \ref{fig:ldos}(c)) the main issue arising from the interaction between the two structures is a $p$-type doping effect due to the presence of the high work function metal (see $\Delta E_F$ in Fig. \ref{fig:ldos}(c) for the difference between the Fermi levels of the metal-contacted and the respective ideal aGNR). Hole carrier injection has been obtained for all high work function metals on metallic GNRs in this study while a less pronounced electron doping effect has been observed in the case of $Al$. Band-bending is also evident here from the first $\pi-\pi^*$ bands and onwards, however the presence of the electrostatic potential does not seem to affect the states that lie inside the first $\pi-\pi^*$ plateau (e.g. see the GNR-long flat line that corresponds to the secondary $meV$ bandgap of the 14 aGNR at the ideal structure's Fermi level in Fig. \ref{fig:ldos}(c)). 

\begin{figure}
	\centering
		\includegraphics[width=0.9\columnwidth]{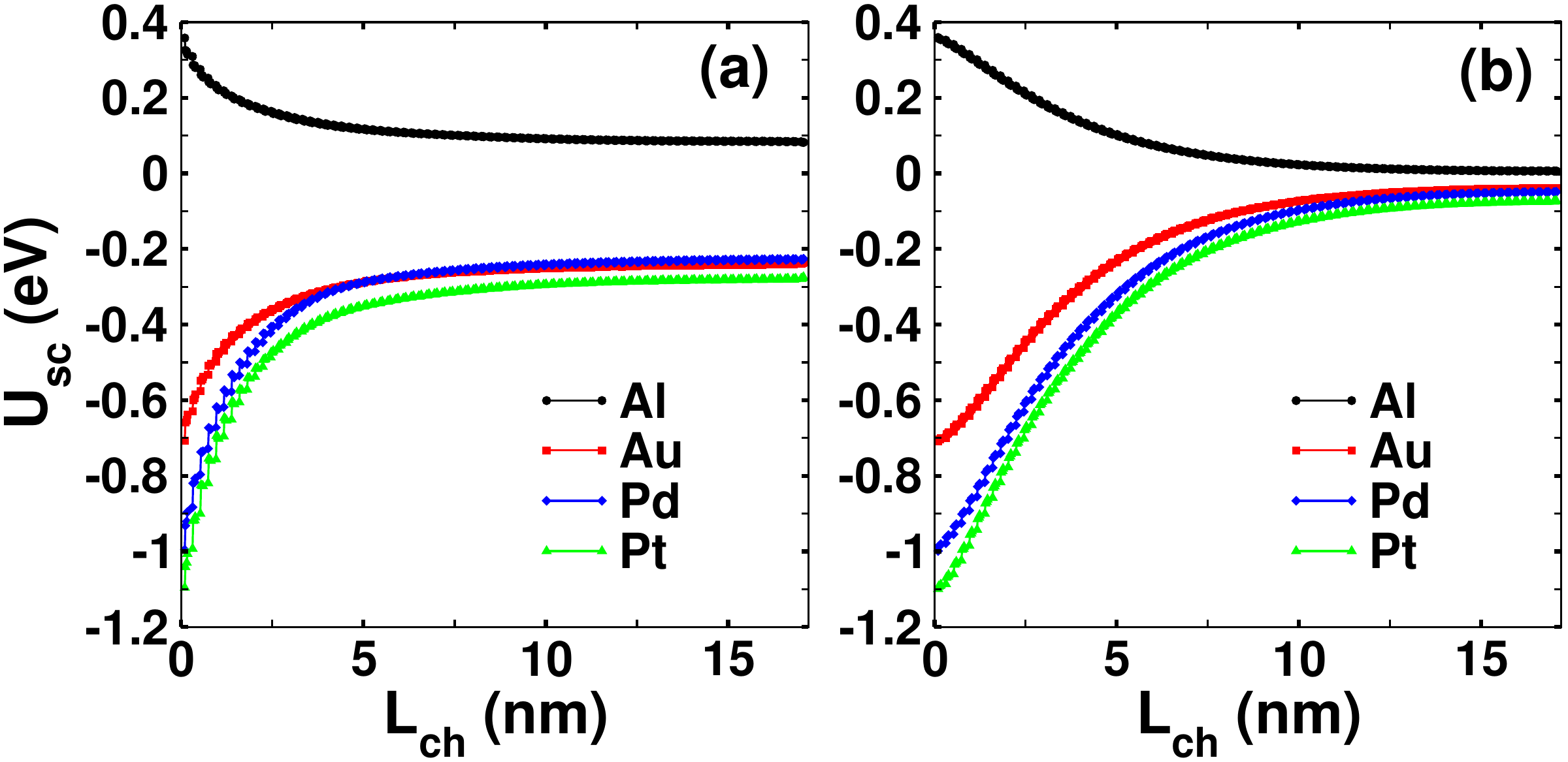}
	\caption{Electrostatic potential profile $U_{sc}$ as a function of the channel length $L_{ch}$ for (a) a $N_a$=14 aGNR and (b) a $N_a$=16 aGNR contacted with $Au$, $Pd$, $Pt$ and $Al$ electrodes.}
	\label{fig:potprof}
\end{figure}

Characteristic 1D junction electrostatics are present in the metal-aGNR case. Fig. \ref{fig:potprof} shows potential profiles along ribbon lengths for the previously shown $N_a$=14 and 16 aGNRs contacted by all available metals in the study. The main aspect of the electrostatic potential for the semimetallic aGNR is a steep potential drop near the contact interface that decays after few $nm$ to a non-zero flat value. This finite potential value denotes the presence of carrier accumulation throughout the GNR length (holes for $Au$, $Pd$ and $Pt$ and electrons for $Al$). In the case of the semiconducting 16 aGNR the Schottky junction behaves qualitatively different. Screening is smoother and charges tend to vanish away from the metal contact, however also in this case long-range depletion tails in the charge distribution have been obtained, in accordance with previous studies on CNT junctions\cite{1999PhRvL..83.5174L}. In this sense an accurate estimation of depletion length scales becomes difficult in these systems and ``breaks'' the traditional metal-semiconductor scheme, giving rise to novel 1D device design possibilities. The categorization of metal-GNR electrostatics on the basis of the conductive character of the respective GNR has been also encountered in the metal-CNT case\cite{2008JChPh.128p4706D}. It can be argued that as the width of semiconducting GNRs grows and the respective bandgaps decrease\cite{2006PhRvL..97u6803S} we can expect an electrostatic response that smoothly shifts from Fig. \ref{fig:potprof}(b) to Fig. \ref{fig:potprof}(a). 

\begin{figure}
	\centering
		\includegraphics[width=0.9\columnwidth]{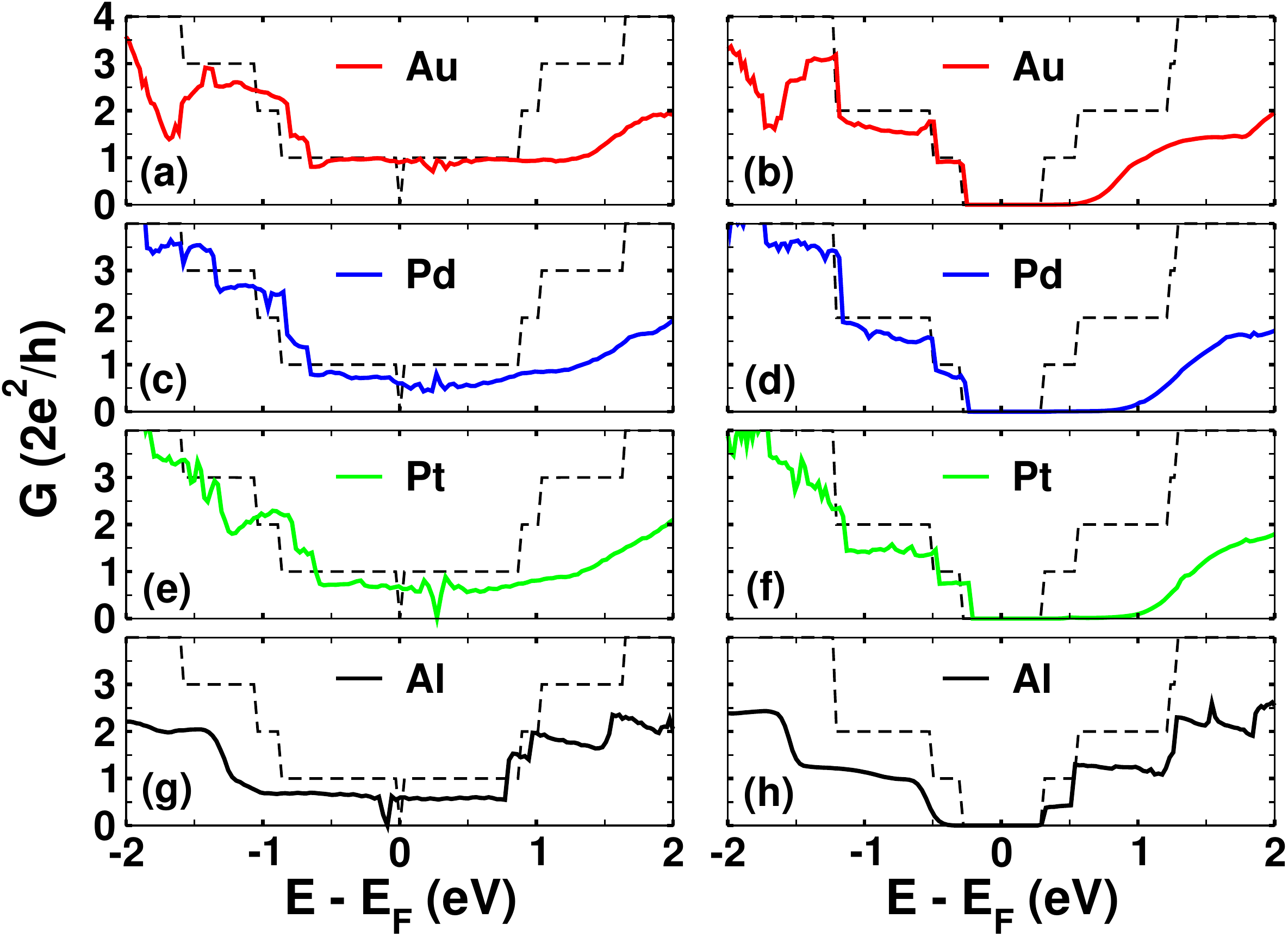}
	\caption{Conductance as a function of energy for a $N_a$=14 aGNR (left column) and a $N_a$=16 aGNR (right column) contacted with: (a-b) $Au$, (c-d)  $Pd$, (e-f)  $Pt$, (g-h)  $Al$. Dashed lines show ideal conductances for the respective aGNRs.}
	\label{fig:cond}
\end{figure}

Fig. \ref{fig:cond} shows the influence of chemical bonding and electrostatics in the conduction mechanism of the studied systems. High work function $Au$, $Pd$ and $Pt$ metals give rise to qualitatively similar transport characteristics that originate from the electrostatic aspect of the heterojunctions. Namely, $p$-type conduction characteristics have been obtained for the 14 aGNR and low Schottky barriers with respect to the valence band (of the order of 0.2-0.3 eV) for the 16 aGNR. Fermi level to conduction band distances increase for the semiconducting ribbon with respect to the ideal case, arriving at $E_C - E_F \sim 1 eV$ for $Pd$ and $Pt$. In all cases conduction band charge flow is strongly suppressed, giving rise to a selective loss of the quantization steps that are typical of the 1D subbands in GNR structures. This behavior is related with the smooth bending of the conduction band that creates a state-free zone near the interface (see Fig. \ref{fig:ldos}(a)). The combination of $p$-type characteristics and conductance suppression due to band-bending gives an asymmetric form to the overall conductance distribution (as similarly calculated also for CNTs\cite{2002PhRvL..89j6801H}). In terms of chemical bonding only $Au$ seems transparent near the Fermi level with the conductance arriving at the $1 G_0$ plateau of the ideal case, whereas $Pd$ and $Pt$ demonstrate a slightly lower transparency. On the other hand, valence band transparency above the first conductance plateau is enhanced for $Pd$ and $Pt$, which show a smaller extent of conductance fluctuations with respect to $Au$, making them more appropriate for high bias electrical measurements. A careful comparison between group 10 transition metals $Pd$ and $Pt$ shows that nonetheless the similarities deriving from their electronic structures, $Pd$ shows a slightly better conductance response in the quantization steps of the valence band. The case of low work function $Al$ electrode is distinct, since despite the contact-induced $n$-type doping (for the 14 aGNR) and quasi-ambipolar Schottky behavior (for the 16 aGNR), the dominant aspect that characterizes conduction is the strong scattering by the contacts. Here contact resistance constitutes the main factor of conductance suppression with respect to the ideal case, with quasi-blocked conduction channels and overall conductance degradation throughout the energy spectrum. It is therefore clear that the electrostatics and chemical bonding act complementary in metal-graphene nanostructures and a categorization of the metallic contacts on the basis of their transparency to graphene should incorporate a best compromise between these two aspects. Finally it should be noted that localized gap states that form near the metal-GNR interface (see Fig. \ref{fig:ldos}(a),(b)) do not contribute to the transport process.


\begin{figure}
	\centering
		\includegraphics[width=0.9\columnwidth]{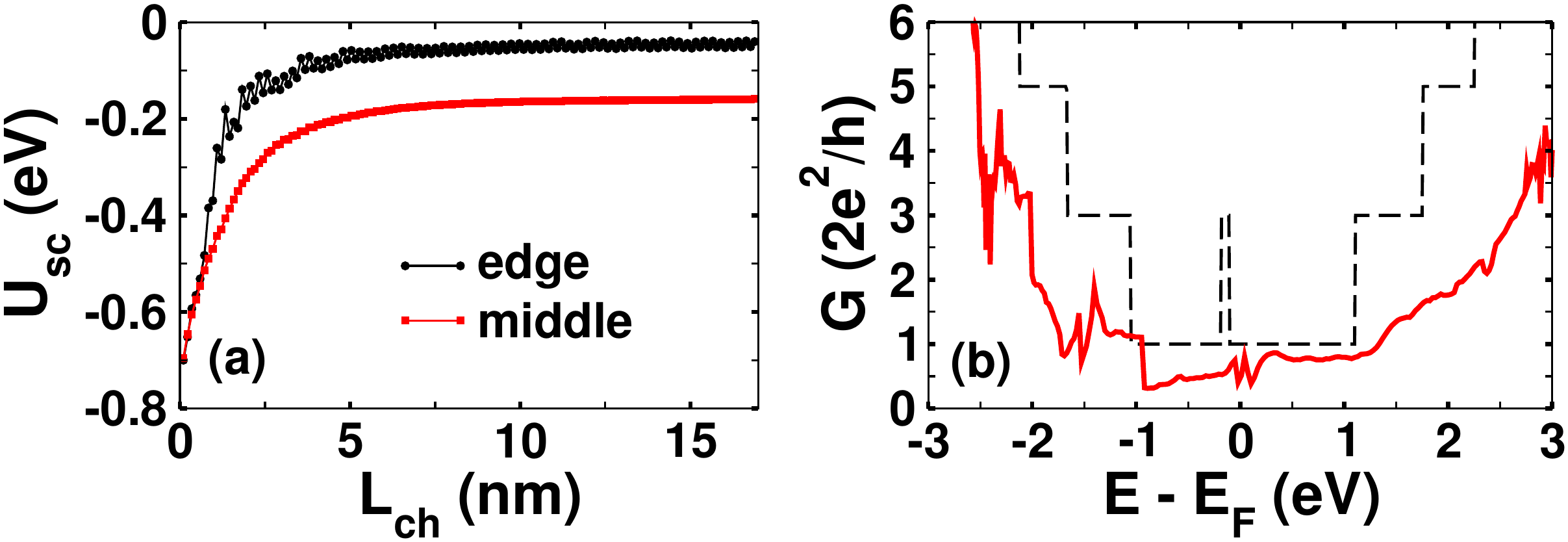}
	\caption{Electrostatic potential profile $U_{sc}$ as a function of the channel length $L_{ch}$ (a) and conductance as a function of energy (b) for a $N_z$=10 zGNR contacted with $Au$. The dashed line shows the ideal conductance of the zGNR.}
	\label{fig:zigzag}
\end{figure}

Junctions between metals and zGNRs preserve similar qualitative characteristics with respect to aGNRs. However, the presence of the edge states in the channel material\cite{2006PhRvL..97u6803S} and the accompanying large DOS near the Fermi level of these systems strongly enhances the role of localized electron-electron interactions. Hence, contrary to aGNRs, the electrostatic response is not uniform throughout the zGNR width and gives rise to a faster potential screening near the borders than in the center of the zGNR (Fig. \ref{fig:zigzag}(a)). Moreover, the reduced area of interface overlap between  metallic and edge wavefunctions further hinders the transparent transmission of electrons in these systems (see the lower conductance with respect to the ideal case in Fig. \ref{fig:zigzag}(b)). It should be noted though that by the suppression of the edge state (e.g. due to corrugation from nanolithographic processes), transport and electrostatic properties are expected to converge towards the aGNR case.

To conclude, this study has addressed the problem of metal-GNR heterojunctions within an atomistic approach that deals with both the electrostatics as well as the chemical aspects of the interface. Results have shown that band-bending, doping and  bonding characteristics of this interaction can non-trivially influence the conduction mechanism, giving rise to conductance asymmetries, Schottky barriers and suppression of ideal transport properties. This study also implies that the electrostatics and the chemical bonding aspects can act complementary for the determination of contact transparency in graphene. GNRs, as 1D $sp^2$ carbon allotrope systems share a lot of common properties with CNTs. Within a certain qualitative framework, this work argues that theoretical/experimental knowledge obtained for metal-CNT heterojunctions can be also valid in the case of GNRs. It is therefore crucial to understand the pros and cons of the two systems in terms of fabrication/growth/patterning methods and electrical/mechanical/optical characteristics in order to distinguish the ideal candidate for post-Si nanoelectronic applications.

I.D. and A.L. acknowledge the European Science Foundation (ESF) under the EUROCORES Programme EuroGRAPHENE CRP GRAPHIC-RF for partial financial support. G.F. and G.I. acknowledge the EC Seventh Framework Program Project GRAND (Contract n. 215752) and the MIUR-PRIN (Contract n. 2008S2CLJ9).

\bibliography{Schottky}

\end{document}